\pdfoutput=1

\documentclass[aps,prd,onecolumn,twoside,notitlepage,preprintnumbers,showpacs,showkeys,superscriptaddress,byrevtex,floatfix,nofootinbib,amsmath,amssymb]
{revtex4}
\usepackage{array,dsfont,graphicx,mathtools,placeins,verbatim,tensor}
\usepackage[bookmarksnumbered,bookmarksopen=true,bookmarksopenlevel=1,pdfstartview=FitH]{hyperref}
\usepackage[all]{hypcap}

\newcolumntype{M}[1]{>{$}{#1}<{$}}

\def\0{{\sst{(0)}}}
\def\1{{\sst{(1)}}}
\def\2{{\sst{(2)}}}
\def\3{{\sst{(3)}}}
\def\4{{\sst{(4)}}}
\def\5{{\sst{(5)}}}
\def\6{{\sst{(6)}}}
\def\7{{\sst{(7)}}}

 \def\bd{\begin{document}} \def\ed{\end{document}}
\def\ds{\documentstyle} \let\fr=\frac \let\bl=\bigl \let\br=\bigr
\let\Br=\Bigr \let\Bl=\Bigl 
\let\bm=\bibitem
\let\na=\nabla
\let\pa=\partial \let\ov=\overline 
\newcommand{\be}{\begin{equation}} 
\newcommand{\ee}{\end{equation}} 
\def\ba{\begin{array}}
\def\ea{\end{array}}
\def\ft#1#2{{\textstyle{{\scriptstyle #1}\over {\scriptstyle #2}}}}
\def\fft#1#2{{#1 \over #2}}
\def\del{\partial}
\def\sst#1{{\scriptscriptstyle #1}}
\def\oneone{\rlap 1\mkern4mu{\rm l}}
\def\ie{{\it i.e.\ }}
\def\via{{\it via}}
\def\semi{{\ltimes}}
\def\v{{\cal V}}

\newcommand{\bp}{\bullet}
\newcommand{\sfx}{\textsf{x}}
\newcommand{\sfo}{\textsf{o}}
\newcommand{\alg}[1]{\ensuremath{\mathfrak{#1}}}
\newcommand{\rep}[1]{\ensuremath{\mathbf{#1}}}
\newcommand{\fld}[1]{\ensuremath{\mathds{#1}}}
\newcommand{\rng}[1]{\ensuremath{\mathds{#1}}}
\newcommand{\SUSY}{\ensuremath{\mathcal{N}}}

\newcommand{\ho}[1]{$\, ^{#1}$}
\newcommand{\hoch}[1]{$\, ^{#1}$}
\newcommand{\bea}{\begin{eqnarray}} 
\newcommand{\eea}{\end{eqnarray}} 
\newcommand{\ra}{\rightarrow}
\newcommand{\lra}{\longrightarrow}
\newcommand{\Lra}{\Leftrightarrow}
\newcommand{\tr}{{\rm tr} }
\newcommand{\Tr}{{\rm Tr} } 

\DeclareMathOperator{\Span}{span}

\begin{document}

\title{Generalized mirror symmetry and trace anomalies}  
\author{M. J. Duff}
\email[]{m.duff@imperial.ac.uk}
\affiliation{Theoretical Physics, Blackett Laboratory, Imperial College London, London SW7 2AZ, United Kingdom}
\author{S. Ferrara}
\email[]{sergio.ferrara@cern.ch}
\affiliation{Physics Department, Theory Unit, CERN, CH 1211, Geneva 23, Switzerland}
\affiliation{INFN - Laboratori Nazionali di Frascati, Via Enrico Fermi 40, 00044 Frascati, Italy}
\affiliation{Department of Physics and Astronomy, University of California, Los Angeles, CA USA}
\date{\today}

\begin{abstract}
We consider the compactification of M-theory on $X^7$ with betti numbers $(b_0, b_1, b_2, b_3, b_3, b_2, b_1, b_0)$ and define a generalized mirror symmetry $(b_0,  b_1, b_2, b_3) \rightarrow (b_0, b_1,  b_2 -\rho/2, b_3+\rho/2)$ under which $\rho \equiv 7b_0-5b_1+3b_2 -b_3$ changes sign.
Generalized self-mirror theories with  $\rho=0$  have massless sectors with vanishing trace anomaly (before dualization).  Examples  include pure supergravity with ${\cal N} \geq 4$ and supergravity plus matter with ${\cal N} \leq 4$.
\end{abstract}


\begin{flushright}
\hfill{Imperial/TP/2009/mjd/3}\\
\hfill{CERN-PH-TH/2010-162}\\
\end{flushright}
\vspace{10pt}
\maketitle
\tableofcontents
\newpage
\FloatBarrier
\section{Introduction}
\label{Introduction}
We consider compactification of $({\cal N}=1,D=11)$ supergravity on a 7-manifold $X^7$ with betti numbers ($b_0,  b_1,  b_2, b_3, b_3, b_2, b_1, b_0$)  and define a generalized mirror symmetry
\begin{equation}
(b_0,  b_1, b_2, b_3) \rightarrow (b_0, b_1,  b_2 -\rho/2, b_3+\rho/2)
\label{mirror}
\end{equation}
under which
\begin{equation}
\rho(X^7) \equiv 7b_0-5b_1+3b_2 -b_3
\end{equation}
changes sign
\begin{equation}
\rho \rightarrow -\rho
\end{equation}
Generalized self-mirror theories are defined to be those for which $\rho$ vanishes.  In the case of $G_2$ manifolds with $b_1=0$,  Joyce \cite{Joyce:1996a,Joyce:1996b} refers to $\rho=0$ as an ``axis of symmetry''.  For related work on mirror symmetry and Joyce-manifiolds, see \cite{Shatashvili:1994zw,Acharya:1997rh,Gaberdiel:2004vx}.

The massless sectors of these compactifications have 
\begin{equation}
f=4(b_0+b_1+b_2+b_3)
\end{equation}
degrees of freedom.  Interestingly enough,  we shall see in section \ref{D=4} that the quantity $\rho$ also shows up  in their on-shell trace anomaly \cite{Duff:1977ay,Duff:1993wm} 
\begin{equation}
g_{\mu\nu}<T^{\mu\nu}>=A \frac{1}{32\pi^2}R^{*}{}^{\mu\nu\rho\sigma}R^{*}{}_{\mu\nu\rho\sigma}
\end{equation}
 which is given by 
\begin{equation}
A=-\frac{1}{24}\rho.
\end{equation}
Hence generalized self-mirror theories have vanishing anomaly with betti numbers\footnote{We assume that there is a $U(1)^{b_1}$ isometry, which will be the case for $X^{(8-{\cal N})}\times T^{({\cal N}-1)}$ with $X^{(8-{\cal N})}$ simply connected.}
\begin{equation}
(b_0, b_1, b_2, b_3)= (1, {\cal N}-1, n, 3n-5{\cal N}+12)
\end{equation}
and
\begin{equation}
f=16(n-{\cal N}+3)
\end{equation}
degrees of freedom, where $1 \leq {\cal N} \leq 8$ is the number of supersymmetries.   If we denote the $D=11$ fields by $(g_{MN};  \psi_{M};  A_{MNP})$ and the corresponding $D=4$ fields by  $(g_{\mu\nu}, {\cal A}_{\mu}, {\cal A};  \psi_{\mu}, \chi; A_{\mu\nu\rho}, A_{\mu\nu}, A_{\mu}, A)$, then the possible generalized self-mirror theories and their betti numbers are:

\begin{itemize}

\item{${\cal N}=1, n \geq 0$, $f=16(n+2)$}

 $X^7: (1, 0, n, 3n+7, 3n+7, n, 0, 1)$ 
 
 yielding 1 graviton $(g_{\mu\nu};\psi_{\mu}; A_{\mu\nu\rho})$ plus $n$ vector $(\chi; A_{\mu})$ plus $(3n+7)$ chiral $({\cal A}; \chi; A)$.

\item{${\cal N}=2, n \geq 0$, $f=16(n+1)$}

$X^6:(1, 0, n, 2n+2, n, 0, 1)$;  $X^6 \times S^1: (1, 1, n, 3n+2, 3n+2, n, 1, 1)$  

 yielding 1 graviton $(g_{\mu\nu}, {\cal A}_{\mu}; 2\psi_{\mu}; A_{\mu\nu\rho})$ plus $n$ vector $({\cal A}; 2\chi; A_{\mu}, A)$ plus $n$ hyper  $(2{\cal A}; 2\chi; 2A)$ plus 1 linear $({\cal A}; 2\chi; A_{\mu\nu}, 2A)$.

 \item{${\cal N}=3, n \geq 3$, $f=16n$}
 
  $X^5: (1, 0, n-1, n-1, 0, 1)$;  $X^5 \times T^2: (1, 2, n, 3n-3, 3n-3, n, 2, 1)$ 
  
  yielding 1 graviton $(g_{\mu\nu}, 2{\cal A}_{\mu};  3\psi_{\mu}, \chi; A_{\mu\nu\rho}, A_{\mu})$ plus $(n-3)$ vector $(3{\cal A}; 4\chi; A_{\mu}, 3A)$ plus $2$ 2-form   $(2{\cal A}; \chi; A_{\mu\nu}, A_{\mu}, 3A)$.

  \item{${\cal N}=4, n\geq 6$, $f=16(n-1)$}
  
  $X^4: (1, 0, n-3, 0, 1)$;  $X^4 \times T^3: (1, 3, n, 3n-8, 3n-8, n, 3, 1)$  
  
  yielding 1 graviton $(g_{\mu\nu}, 3 {\cal A}_{\mu},  {\cal A}, 4 \psi_{\mu}, 4 \chi, A_{\mu\nu\rho}, 3  A_{\mu}, A)$ plus $(n - 6)$ vector $(3{\cal A}, 4\chi; A_{\mu}, 3A)$ plus $3$ 2-form $(2{\cal A};  4 \chi; A_{\mu\nu},A_{\mu}, 3 A)$.  
  
  The case $n=25$ corresponds to $X^4=K3$ \cite{Duff:1983vj}.
  
  \item{${\cal N}=5, n=6$, $f=64$}
  
  $X^3: (1, 0, 0, 1)$; $X^3 \times T^4: (1, 4, 6, 5, 5, 6, 4, 1)$   
  
  yielding 1 graviton $(g_{\mu\nu}, 4{\cal A}_{\mu}, {\cal A}; 5\psi_{\mu}, 11\chi; A_{\mu\nu\rho}, 4A_{\mu\nu}, 6A_{\mu}, 5A)$.
  
   \item{${\cal N}=6, n=11$, $f=128$}
  
   $X^2: (1, 0, 1)$; $X^2 \times T^5: (1, 5, 11,15,15, 11, 5, 1)$  
   
   yielding 1 graviton $(g_{\mu\nu}, 5{\cal A}_{\mu}, 10{\cal A}; 6\psi_{\mu}, 16\chi; A_{\mu\nu\rho}, 5A_{\mu\nu},11A_{\mu},15A)$.
   
  \item{${\cal N}=8, n=21$, $f=256$}
  
   $T^7: (1, 7, 21, 35, 35, 21, 7, 1)$ 
   
 yielding 1 graviton $(g_{\mu\nu}, 7{\cal A}_{\mu}, 28 {\cal A}; 8\psi_{\mu}, 56\chi; A_{\mu\nu\rho}, 7A_{\mu\nu},21 A_{\mu},35 A)$.
 \end{itemize}
In listing these results, we simply record what the betti numbers of the compactifying 7-manifold would have to be, without attempting to prove in all cases that such manifolds actually exist.   Of particular interest are the four cases
\begin{equation} 
(b_0,b_1,b_2,b_3)=(1,{\cal N}-1,3{\cal N}-3, 4{\cal N}+3)
\end{equation}
with ${\cal N}=1,2,4,8$, namely the four ``curious'' supergravities, discussed in  
\cite{Duff:2010b}:
$({\cal N}=1, n=0, f=32)$,  $({\cal N}=2, n=3, f=64)$,  $({\cal N}=4, n=9, f=128)$,  $({\cal N}=8, n=21 ,f=256)$, which enjoy some remarkable properties\footnote{ The ${\cal N}=8, 4, 2, 1$ cases are  related  \cite{Cremmer:1979up,Ferrara:1989nm,Sen:1995ff,Gaberdiel:2004vx}  to the orbifolds $T^7$,  ${T^7}/{Z_2}$, ${T^7}/({Z_2 \times Z_2})$ ,   ${T^7}/(Z_2 \times Z_2 \times Z_2)$.}.

In section \ref{IIA} we note that  the case of M-theory on $X^6 \times S^1$ with betti numbers $(b_0, b_1, b_2, b_3, b_3, b_2, b_1, b_0)$  is equivalent to Type IIA on $X^6$ with betti numbers $(c_0, c_1, c_2, c_3, c_2, c_1, c_0)$  related by
\begin{equation}
(b_0 ,b_1, b_2, b_3)= (c_0, c_0+c_1, c_1+c_2,c_2+c_3)
\end{equation}
and hence
\be
\rho(X^6 \times S^1)= \chi(X^6)
\ee
 where $ \chi(X^6)$   is the Euler number of $X^6$  
 \be
 \chi(X^6)=2c_0-2c_1+2c_2 -c_3.
 \ee
 The generalized mirror symmetry transformation  (\ref{mirror}) then becomes 
 \be
 (c_0, c_1, c_2, c_3) \rightarrow  (c_0, c_1, c_2-\chi/2, c_3+\chi)
 \ee
 under which $\chi$ also changes sign
 \begin{equation}
\chi \rightarrow -\chi.
\end{equation}
Further specializing to $X^6$=Calabi-Yau with  betti numbers:  $(1, 0, h^{11},  2+2h^{21}, h^{11}, 0, 1)$ our generalized mirror symmetry reduces to the familiar interchange of hodge numbers $h^{11} \leftrightarrow h^{21}$ \cite{Shing:492595}.  As for the trace anomaly,
\be
A=-\frac{\chi}{24}
\ee
and so in Euclidean signature
\begin{equation}
\int d^4x \sqrt{g}g_{\mu\nu}<T^{\mu\nu}>=-\frac{1}{24}\chi(M^4)\chi(X^6)=-\frac{1}{24}\chi(M^4 \times X^6)
\end{equation}
where $\chi(M^4)$ is the Euler number of spacetime.  

The compactifications of $({\cal N}=1, D=10)$ supergravity on $X^6$ are just given by the NS sector of Type IIA. Their massless sectors have 
\begin{equation}
f=4(2c_0+2c_1+2c_2+c_3)
\end{equation}
degrees of freedom. Their anomaly is given by
\begin{equation}
A=-\frac{1}{24}(65c_0-15c_1+c_2+c_3/2)
\end{equation}
which vanishes when 
\begin{equation}
(c_0, c_1, c_2, c_3)= (1, 2{\cal N}-2, n, 30{\cal N}-95-n)
\end{equation}
and
\begin{equation}
f=4(26{\cal N}-97+3n).
\end{equation}
So the only possibility is:   
  \begin{itemize}
  \item{${\cal N}=4$, $n=15$, $f=128$}
 
    1 graviton $(g_{\mu\nu}, 3 {\cal A}_{\mu},  \Phi, 4 \psi_{\mu}, 4 \chi, A_{\mu\nu}, 3  A_{\mu})$ plus $3$ vector  $(3{\cal A};  4 \chi; A_{\mu}, 3 A)$ + 3 vector $({\cal A}_{\mu}, 4{\cal A};  4 \chi;  2 A)$

 \end{itemize}
 Note that the field content of the ${\cal N}=4$ graviton and vector multiplets arising from compactification of  $({\cal N}=1,D=10)$  on $T^6$ is different from those arising from  $({\cal N}=2,D=10)$ on $X^4 \times T^2$ with $X^4$ betti numbers $(1,0,6,0,1)$.   In particular the anomalies of each multiplet vanish separately. These two versions of  ${\cal N}=4$ are the dual pair discussed in \cite{Sen:1995ff}. Note also that the $({\cal N}=1,D=10)$ vector multiplet $(A_M; \chi)$ appearing in the heterotic string yields the vector $({ A_{\mu}};  4 \chi; 6A)$ on $T^6$ which separately has $A=0$ also.

 In obtaining these results,  we adopt the interpretation of \cite{Duff:1980qv}  that assigns different anomalies to $A_{\mu\nu}$ and ${\cal A}$ even though they are naively dual\footnote{As may be seen even in the apparently simple example of abelian 1-forms in four dimensions, these dualities are quite subtle \cite{Witten:1995gf}.}  to one another  (each with $f=1$ ) and nonzero anomaly to $A_{\mu\nu\rho}$ (with $f=0$).  This is controversial, with some authors agreeing \cite{Nicolai:1980td} and others taking the view that $A_{\mu\nu}$ has the same anomaly as ${\cal A}$ and that $A_{\mu\nu\rho}$ has vanishing anomaly \cite{Siegel:1980ax,Fradkin:1983tg,Grisaru:1984vk,Buchbinder:2008jf}. For the purposes of comparison, we give the results that the alternative view would yield in appendix \ref{dual}.  In particular, for M on $X^7$ and Type IIA on $X^6$
one finds
\begin{equation}
A(M)=-\frac{1}{24}{(41b_0-19b_1-3b_2+b_3)}
\end{equation}
\begin{equation}
A(IIA)=-\frac{1}{24}{(-22c_0+22c_1+2c_2-c_3)}
\end{equation}
\begin{equation}
A(IIB)=\frac{1}{24}(26c_0-26c_1+2c_2-c_3)
\end{equation}
none of which seems to have any separate topological significance (although $A(IIA)-A(IIB)=-{\chi}/{12}$).  All yield a nonzero result  for ${\cal N} > 3$. It should be noted, moreover, that our interpretation is supported by string calculations \cite{Antoniadis:1992sa}.

Given the relation between trace anomalies and logarithmic corrections to black hole entropy 
\cite{Fursaev:1994te,Solodukhin:2008dh,Banerjee:2010qc,Aros:2010jb,Dowker:2010bu}, one is tempted to conclude that these are absent in generalized self-mirror theories. The authors of \cite{Banerjee:2010qc}, however,  do not reach this conclusion and it seems that there are still some unresolved issues.

Finally, in section \ref{Fermion} we introduce a fermionic mirror map
\begin{equation}
(b_0,  b_1, b_2, b_3) \rightarrow (b_0, b_1+{\cal N'}-{\cal N},  b_2 -{\cal N'}+{\cal N}, b_3)
\label{eq:mirrorfermi}
\end{equation}
which preserves the number of spin 2, spin 1 and spin 0 but changes the number of spin 3/2 (from  ${\cal N} $ to ${\cal N}'$) and spin 1/2, keeping $f$ fixed.  Previously known examples \cite{Ferrara:2006yb,Ferrara:2008ap,Roest:2009sn} of  fermionic mirror pairs are provided by $({\cal N}=6,{\cal N}'=2)$, $({\cal N}=4,{\cal N}'=2)$ and  $({\cal N}=3,{\cal N}'=2)$ supergravity plus matter theories. Both members of a pair have exactly the same bosonic field content including interactions. Curiously, the partner with the higher supersymmetry is generalized self-mirror in the bosonic sense. In addition,  we  find a new two-parameter family with $({\cal N}=1,{\cal N}'=2)$.

\FloatBarrier
\section{M on $X^7$}
\label{D=4}

\subsection{Betti numbers}

Consider $X^{(8-{\cal N})}\times T^{({\cal N}-1)}$ compactification of $D=11$ supergravity with 128+128 degrees of freedom
\[
(g_{MN}, \Psi_{M}, A_{MNP})
\]
as shown in Table \ref{D=11}.  We denote  the betti numbers of $X^7$, $X^6$,  $X^5$, $X^4$, $X^3$, $X^2$ by the letters $b$, $c$, $e$, $d$, $j$, $k$, respectively.  The betti numbers of $S^1$ are $(1,1)$,  of $T^2$ are $(1,2,1)$, of $T^3$ are $(1,3,3,1)$, of $T^4$ are $(1,4,6,4,1)$, of $T^5$ are $(1,5,10,10,5,1)$ of $T^7$ are $(1,7,21,35,21,7,1)$, so

\begin{equation}\label{bet2}
\begin{split}
X^7&:(b_0 ,b_1, b_2, b_3)\\
X^6 \times S^1&: (c_0, c_0+c_1, c_1+c_2,c_2+c_3)\\
X^5 \times T^2&: (e_0, 2e_0+e_1, e_0+2 e_1+e_2, e_1+3e_2)\\
X^4 \times T^3&: (d_0, 3d_0+d_1, 3d_0+3d_1+d_2, d_0+4d_1+3d_2)\\
X^3 \times T^4&: (j_0, 4j_0+j_1, 7j_0+4j_1, 5j_0+10j_1)\\
X^2 \times T^5&: (k_0, 5k_0+k_1, 11k_0+5k_1, 15k_0+10k_1).
\end{split}
\end{equation}

\FloatBarrier
\subsection{Trace anomalies}
\label{trace} 
 
The fields in the massless sector of each $D=4$  theory will exhibit an on-shell Weyl anomaly \cite{Duff:1977ay,Duff:1993wm}
\begin{equation}
g_{\mu\nu}<T^{\mu\nu}>=A \frac{1}{32\pi^2}R^{*}{}^{\mu\nu\rho\sigma}R^{*}{}_{\mu\nu\rho\sigma}
\end{equation}
so that in Euclidean signature
\begin{equation}
\int d^4x \sqrt{g}g_{\mu\nu}<T^{\mu\nu}>=A \chi(M^4)
\end{equation}
where $\chi(M^4)$ is the Euler number of spacetime.  We adopt the interpretation of \cite{Duff:1980qv}  that assigns different anomalies to $A_{\mu\nu}$ and ${\cal A}$ even though they are naively dual to one another  (each with one degree of freedom) and nonzero anomaly to $A_{\mu\nu\rho}$ (with zero degrees of freedom). Starting with a Lagrangian
\be
-\frac{1}{2}\phi \Delta \phi
\ee
the one-loop effective action is
\be
ln ({\rm det}~ \Delta)^{-1/2}.
\ee
The total trace of the stress tensor, which we refer to as the ``anomaly'' even when $\Delta$ is not conformal,  is then given by the Schwinger-De Witt coefficients $B$, which in four spacetime dimensions are quadratic in the curvature. When the operator is the laplacian on $p$-forms $\Delta_p$, the corresponding coefficients $B_p$ obey
\be
\int d^4x (B_0-B_1+B_2-B_3+B_4) =\frac{1}{32\pi^2}\int d^4x *R*R=\chi(M^4)= b_0-b_1+b_2-b_3+b_4
\ee
The ghost-for-ghost procedure \cite{Siegel:1980jj} means that we have
\begin{equation}\label{bet1}
\begin{split}
p&=0: B_0\\
p&=1: B_1-2B_0\\
p&=2: B_2-2B_1+3B_0\\
p&=3: B_3-2B_2+3B_1-4B_0\\
p&=4: B_4-2B_3+3B_2-4B_1+5B_0\\
\end{split}
\end{equation}
Bearing in mind  $B_p=B_{(4-p)}$, we find
\be
A_2-A_0=1
\ee
even though both describe one degree of freedom
and
\be
A_3=-2
\ee
\be
A_4=3
\ee
even though both describe zero degrees of freedom. In fact for $p \geq 3$
\be
A_p=(-1)^p(p-1).
\ee

The value of the $A$ coefficient for each supergravity field \cite{Duff:1977ay,Duff:1993wm}  is given in Table \ref{D=11}.  Remarkably, we find that the total anomaly depends on $\rho$
\begin{equation}
A=-\frac{1}{24}\rho(X^7).
\end{equation}
So the anomaly flips sign under generalized mirror symmetry and vanishes for generalized self-mirror theories. In the case of  $({\cal N}=1,D=11)$ on $X^6 \times S^1$, or equivalently  (Type IIA, $D=10$) on $X^6$,
\begin{equation}
A=-\frac{1}{24}\chi(X^6)
\end{equation}
and so in Euclidean signature
\begin{equation}
\int d^4x \sqrt{g}g_{\mu\nu}<T^{\mu\nu}>=-\frac{1}{24}\chi(M^4)\chi(X^6)=-\frac{1}{24}\chi(M^4 \times X^6)
\end{equation}
where $\chi(M^4)$ is the Euler number of spacetime.  It would be interesting to see if this formula generalizes to other spacetime dimensions.

For $X^{(8-{\cal N})}\times T^{({\cal N}-1)}$ with ${\cal N} \geq 3$ the anomaly vanishes identically as shown in Table \ref{D=11}.   Of particular interest are the four cases
\begin{equation} 
(b_0,b_1,b_2,b_3)=(1,{\cal N}-1,3{\cal N}-3, 4{\cal N}+3)
\end{equation}
with ${\cal N}=1,2,4,8$, namely the four ``curious'' supergravities, discussed in  
\cite{Duff:2010b}:
$({\cal N}=1, n=0, f=32)$,  $({\cal N}=2, n=3, f=64)$,  $({\cal N}=4, n=9, f=128)$,  $({\cal N}=8, n=21 ,f=256)$, which enjoy some remarkable properties.
\begin{table}[h!]
$\begin{array}{llrrrrrrrrrrrrrrr}
&Field &f&360A&X^7& X^6 \times S^1&X^5 \times T^2&X^4 \times T^3 &X^3 \times T^4&X^2 \times T^5&T^7 \\
&&&&&&&\\
\bigskip
&&&&\\
g_{MN}&g_{\mu\nu}&2&848&b_0&c_0&e_0&d_0&j_0&k_0&1\\
~&{\cal A}_{\mu}&2&-52&b_1&c_0+c_1&2e_0+e_1&3d_0+d_1&4j_0+j_1&5k_0+k_1&7\\
~&{\cal A}&1&4&-b_1+b_3 &-c_0-c_1+c_2+c_3&-2e_0+3e_2&-2d_0+3d_1+3d_2&j_0+9j_1&10k_0+9k_1&28\\
\psi_{M}&\psi_{\mu}&2&-233&b_0+b_1&2c_0+c_1&3e_0+e_1&4d_0+d_1&5j_0+j_1&6k_0+k_1&8\\
~&\chi&2&7&b_2+b_3&c_1+2c_2+c_3&e_0+3e_1+4e_2&4d_0+7d_1+4d_2&11j_0+15j_1&26k_0+15k_1&56\\
A_{MNP}&A_{\mu\nu\rho}&0&-720&b_0&c_0&e_0&d_0&j_0&k_0&1\\
~&A_{\mu\nu}&2&364&b_1&c_0+c_1&2e_0+e_1&3d_0+d_1&4j_0+j_1&5k_0+k_1&7\\
~&A_{\mu}&2&-52&b_2&c_1+c_2&e_0+2e_1+e_2&3d_0+3d_1+d_2&6j_0+5j_1&11k_0+5k_1&21\\
~&A&1&4&b_3&c_2+c_3&e_1+3e_2&d_0+4d_1+3d_2&5j_0+10j_1&15k_0+5k_1&35\\
&&&&&\\
&&&&&&&\\
total ~A&&&&-\rho/24&-\chi/24&0&0&0&0&0&\\
\end{array}$
\caption{ $X^{(8-{\cal N})}\times T^{({\cal N}-1)}$ compactification of $D=11$ supergravity}
\label{D=11}
\end{table}

\FloatBarrier
\subsection{Multiplets}
Here we group the individual fields into supermultiplets as shown in Tables \ref{Y7} to \ref{Y0}, making use of table \ref{4}.

\begin{table}[h!]
$\begin{array}{lrrrrrrrrrrrrrrrr}
Field 			&360A 			&{\cal N}=8~ graviton 		&&{\cal N}=4 ~graviton&{\cal N}=4~gravitino		&{\cal N}=4 ~vector_A & {\cal N}=4~ vector_{\cal A}\\
\bigskip
&&&&&\\
g_{\mu\nu}&848&{\bf 1}&&{\bf 1}&\\
{\cal A}_{\mu}&-52&{\bf 7}&&{\bf 3}&4.{\bf 1}&&\\
{\cal A}&4&{\bf 27}&&&4.{\bf 3}&9.{\bf 1}&{\bf 1}+{\bf 5} \\
\Phi&4&{\bf 1}&&{\bf 1}&\\
\psi_{\mu}&-233&{\bf 8}&&2.{\bf 2}&2.{\bf 2}\\
\chi&7&{\bf 8+48}&&2.{\bf 2}&10.{\bf 2}+2.{\bf 4}&6.{\bf 2}&2.{\bf 2}+2.{\bf 4}&\\
A_{\mu\nu\rho}&-720&{\bf 1}&&{\bf 1}\\
A_{\mu\nu}&364&{\bf 7}&&&4.{\bf 1}&&{\bf 3}\\
A_{\mu}&-52&{\bf 21}&&3.{\bf 1}&4.{\bf 3}&3.{\bf 1}&{\bf 3}\\
A&4&{\bf 35}&&{\bf 1}&4.{\bf 1}+4.{\bf 3}&3.{\bf 3}&3.{\bf 3}\\
&&&&&&&\\
{}~&&A=0&&A=3&A=0&A=0&A=-3\\
\end{array}$
\caption{  $({\cal N}=8,SO(7)) \rightarrow ({\cal N}=4,SO(3)) $  decomposition appropriate for M and Type IIA compactifications. }
\label{4}
\end{table}

\begin{table}[h!]
$\begin{array}{llrrrrrrrrrr}
{\cal N}=1~&multiplet  &f&360A&{\cal N}=b_0+b_1 &  {\cal N}=8 &  {\cal N}=1 & \\
&&&&&&&\\
\bigskip
graviton&(g_{\mu\nu};\psi_{\mu}; A_{\mu\nu\rho})&2+2&-105&b_0&1&1\\
gravitino&({\cal A}_{\mu}; \psi_{\mu}) &2+2&-285&b_1&7&0 \\
&&&&&&&\\
vector&(\chi; A_{\mu})&2+2&-45&b_2&21&n\\
&&&&&&&\\
 chiral &({\cal A}; \chi; A)&2+2&15&-b_1+b_3&28&3n+7\\
&&&&&&\\
linear&(\chi; A_{\mu\nu}, A)&2+2&375&b_1&7&0\\
&&&&&&\\
total ~f &&&&4(b_0+b_1+b_2+b_3)&256&16(n+2)\\
&&&&&&\\
total ~A &&&&-(7b_0-5b_1+3b_2-b_3)/24&0&0\\
\end{array}$
\caption{ The $D=4$ multiplets in an  ${\cal N}$=1 basis. }
\label{Y7}
\end{table}

\begin{table}[h!]
$\begin{array}{llrrrrrrrrrrrrr}
{\cal N}=2~& multiplet  &f&360A&{\cal N}=2c_0+c_1 &  {\cal N}=8 &  {\cal N}=2 & \\
&&&&&&&\\
\bigskip
graviton&(g_{\mu\nu}, {\cal A}_{\mu}; 2\psi_{\mu}; A_{\mu\nu\rho})&4+4&-390&c_0&1&1\\
gravitino&({\cal A}_{\mu}; \psi_{\mu}, \chi; A_{\mu}) &4+4&-330&c_1&6&0 \\
&&&&&&&\\
vector&({\cal A}, 2\chi; A_{\mu}, A)&4+4&-30&c_2&15&n\\
&&&&&&&\\
hyper&(2{\cal A}; 2\chi; 2A)&4+4&30&-c_0-c_1+c_3/2&3&n\\
&&&&&&\\
linear&({\cal A}; 2\chi; A_{\mu\nu}, 2A)&4+4&390&c_0+c_1&7&1\\
&&&&&&\\
total ~f &&&&4(2c_0+2c_1+2c_2+c_3)&256&16(n+1)\\
&&&&&&\\
total ~A &&&&-(2c_0-2c_1+2c_2-c_3)/24&0&0\\
\end{array}$
\caption{  The $D=4$ multiplets in an  ${\cal N}$=2 basis.}
\label{Y6}
\end{table}

\begin{table}[h!]
$\begin{array}{llrrrrrrrrrr}
{\cal N}=3~& multiplet  &f&360A&{\cal N}=3e_0+e_1 &  {\cal N}=8 & {\cal N}=3 & \\
&&&&&&&\\
\bigskip
graviton&(g_{\mu\nu}, 2{\cal A}_{\mu};  3\psi_{\mu}, \chi; A_{\mu\nu\rho}, A_{\mu})&8+8&-720&e_0&1&1\\
gravitino&({\cal A}_{\mu}, {\cal A}; \psi_{\mu}, 3\chi;  2A_{\mu}, A) &8+8&-360&e_1&5&0 \\
&&&&&&&\\
vector&(3{\cal A}; 4\chi; A_{\mu}, 3A)&8+8&0&-2e_0-e_1+e_2&3&n-3\\
&&&&&&\\
2-form&(2{\cal A}; 4\chi; A_{\mu\nu}, A_{\mu}, 3A)&8+8&360&2e_0+e_1&7&2\\
&&&&&&&\\
total ~f &&&&16(e_0+e_1+e_2)&256&16n\\
&&&&&&\\
total ~A &&&&0&0&0\\
\end{array}$
\caption{   The $D=4$ multiplets in an  ${\cal N}$=3 basis.}
\label{Y5}
\end{table}

\begin{table}[h!]
$\begin{array}{llrrrrrrrrrr}
{\cal N}=4~  & multiplet&f&360A& {\cal N}=4d_0+d_1 &  {\cal N}=8 &  {\cal N}=4 & \\
&&&&&&&\\
\bigskip
graviton&(g_{\mu\nu}, 3 {\cal A}_{\mu},  {\cal A}, 4 \psi_{\mu}, 4 \chi, A_{\mu\nu\rho}, 3  A_{\mu}, A)&16+16&-1080&d_0&1&1\\
gravitino&({\cal A}_{\mu},  3{\cal A},  \psi_{\mu}, 7 \chi,  A_{\mu\nu}, 3A_{\mu},4A) &16+16&0&d_1&4&0 \\
&&&&&&&\\
vector&(3{\cal A};  4 \chi; A_{\mu}, 3 A)&8+8&0&-3d_0+d_2&3&n-6\\
&&&&&\\
2-form &(2{\cal A};  4 \chi; A_{\mu\nu},A_{\mu}, 3 A)&8+8&360&3d_0&3&3\\
&&&&&&&\\
total ~f &&&&16(2d_0+2d_1+d_2)&256&16(n-1)\\
&&&&&&\\
total ~A &&&&0&0&0\\
\end{array}$
\caption{  The $D=4$ multiplets in an  ${\cal N}$=4 basis.}
\label{Y4}
\end{table}

\begin{table}[h!]
$\begin{array}{llrrrrrrrrrr}
{\cal N}=5~  & multiplet&f&360A&{\cal N}=5j_0+j_1&{\cal N}=8&{\cal N}=5  \\
&&&&&&&\\
\bigskip
graviton&(g_{\mu\nu}, 4{\cal A}_{\mu}, {\cal A}; 5\psi_{\mu}, 11\chi; A_{\mu\nu\rho}, 4A_{\mu\nu}, 6A_{\mu}, 5A)
&32+32&0&j_0&1&1\\
gravitino&({\cal A}_{\mu},  9{\cal A},  \psi_{\mu}, 15 \chi,  A_{\mu\nu}, 5A_{\mu}, 10A) &32+32&0&j_1&3&0 \\
&&&&&&&\\
total ~f &&&&64(j_0+j_1)&256&64\\
&&&&&&&\\
total ~A &&&&0&0&0\\
&&&&&&\\
\end{array}$
\caption{ The $D=4$ multiplets in an  ${\cal N}$=5 basis.}
\label{Y3}
\end{table}

\begin{table}[h!]
$\begin{array}{llrrrrrrrrrr}
{\cal N}=6~  & multiplet&f&360A&{\cal N}=6k_0+k_1&{\cal N}=8&{\cal N}=6  \\
&&&&&&&\\
\bigskip
graviton&(g_{\mu\nu}, 5{\cal A}_{\mu}, 10{\cal A}; 6\psi_{\mu}, 26\chi; A_{\mu\nu\rho}, 5A_{\mu\nu},11A_{\mu},15A)
&64+64&0&k_0&1&1\\
gravitino&({\cal A}_{\mu},  9{\cal A},  \psi_{\mu}, 15\chi,  A_{\mu\nu}, 5A_{\mu},10A) &32+32&0&k_1&2&0 \\
&&&&&&&\\
total ~f &&&&64(2k_0+k_1)&256&128\\
&&&&&&&\\
total ~A &&&&0&0&0\\
\end{array}$
\caption{  The $D=4$ multiplets in an  ${\cal N}$=6 basis.}
\label{Y2}
\end{table}

\begin{table}[h!]
$\begin{array}{llrrrrrrrrrr}
{\cal N}=8~  & multiplet&f&360A&{\cal N}=8  \\
&&&&&&&\\
\bigskip
graviton&(g_{\mu\nu}, 7{\cal A}_{\mu}, 28 {\cal A}; 8\psi_{\mu}, 56\chi; A_{\mu\nu\rho}, 7A_{\mu\nu},21 A_{\mu},35 A)
&256&0&1&\\
&&&&&&&\\  
total ~f &&&&256\\
&&&&&&&\\
total ~A &&&&0&\\
\end{array}$
\caption{  The $D=4$ multiplets in an  ${\cal N}$=8 basis.}
\label{Y0}
\end{table}


\FloatBarrier
\section{From $D=10$  on $X^6$}
\label{IIA}
\subsection{IIA}

Consider Type IIA in $D=10$.  In the NS sector we have the fields
$(g_{MN}, \Phi; \psi_M, \chi; {A}_{MN})$ with $f=64+64$; in the RR sector we have the fields $({\cal A}_M;\psi_M, \chi; A_{MNP})$ also with $f=64+64$.  We compactify on generic $X^6$ with independent betti numbers $(c_0,c_1,c_2,c_3)$ and on $T^6$ with $(1,6.15.20)$. The results for $NS$ and $RR$ separately and combined are shown in Table \ref{D=10A}.

\begin{table}[h!]
$\begin{array}{llrrrrrrrrrrrrrrr}
Field &f&360A&NS&T^6&RR&T^6&IIA&T^6\\
&&&&&&&\\
\bigskip
&&&&\\
g_{\mu\nu}&2&848&c_0&1&0&0&c_0&1\\
{\cal A}_{\mu}&2&-52&c_1&6&c_0&1&c_0+c_1&7\\
{\cal A}&1&4&-2c_0-2c_1+c_2+c_3&21&c_1&6&-2c_0-c_1+c_2+c_3&27\\
\Phi&1&4&c_0&1&0&0&c_0&1\\
\psi_{\mu}&2&-233&c_0+c_1/2&4&c_0+c_1/2&4&2c_0+c_1&8\\
\chi&2&7&c_1/2+c_2+c_3/2&28&c_1/2+c_2+c_3/2&28&c_1+2c_2+c_3&56\\
A_{\mu\nu\rho}&0&-720&0&0&c_0&1&c_0&1\\
A_{\mu\nu}&1&364&c_0&1&c_1&6&c_0+c_1&7\\
A_{\mu}&2&-52&c_1&6&c_2&15&c_1+c_2&21\\
A&1&4&c_2&15&c_3&20&c_2+c_3&35\\
&&&&&&&\\
total~f&&&2(2c_0+2c_1+2c_2+c_3)&128&2(2c_0+2c_1+2c_2+c_3)&128&4(2c_0+2c_1+2c_2+c_3)&256\\
total~A&&&(65c_0-15c_1+c_2+c_3/2)/24&0&(-67c_0+17c_1-3c_2+c_3/2)/24&0&-(2c_0-2c_1+2c_2-c_3)/24&0\\
\end{array}$
\caption{ $X^6$ compactification of $D=10$ Type IIA sugravity}
\label{D=10A}
\end{table}
\FloatBarrier
From Table \ref{D=10A} we have
\begin{equation}
\begin{split}
A(NS)&=\frac{1}{24}(65c_0-15c_1+c_2+c_3/2)\\
A(RR)&=\frac{1}{24}(-67c_0+17c_1-3c_2+c_3/2)\\
A(IIA)&=-\frac{1}{24}(2c_0-2c_1+2c_2-c_3)=-\frac{1}{24}\chi\\
\end{split}
\end{equation}

Now consider Type IIA on ${\tilde X}^6$, the mirror of $X^6$, with betti numbers
\be
(c_0,c_1,-c_0+c_1+c_3/2, 2c_0+2c_1+2c_2)
\ee
The NS sector remains unchanged and from Table \ref{D=10A'} we have
\begin{equation}
\begin{split}
\tilde A({ {NS}})&=\frac{1}{24}(65c_0-15c_1+c_2+c_3/2)\\
\tilde A({{RR}})&=\frac{1}{24}(-63c_0+13c_1+c_2-3c_3/2)/24\\
\tilde A({ {IIA}})&=\frac{1}{24}(2c_0-2c_1+2c_2-c_3)=\frac{1}{24}\chi\\
\end{split}
\end{equation}

\begin{table}[h!]
$\begin{array}{llrrrrrrrrrrrrrrr}
Field &f&360A&NS&T^6&RR&T^6&IIA&T^6\\
&&&&&&&\\
\bigskip
&&&&\\
g_{\mu\nu}&2&848&c_0&1&0&0&c_0&1\\
{\cal A}_{\mu}&2&-52&c_1&6&c_0&1&c_0+c_1&7\\
{\cal A}&1&4&-c_0-3c_1+2c_2+c_3/2&21&c_1&6&-c_0-2c_1+2c_2+c_3/2&27\\
\Phi&1&4&c_0&1&0&0&c_0&1\\
\psi_{\mu}&2&-233&c_0+c_1/2&4&c_0+c_1/2&4&2c_0+c_1&8\\
\chi&2&7&c_1/2+c_2+c_3/2&28&c_1/2+c_2+c_3/2&28&c_1+2c_2+c_3&56\\
A_{\mu\nu\rho}&0&-720&0&0&c_0&1&c_0&1\\
A_{\mu\nu}&1&364&c_0&1&c_1&6&c_0+c_1&7\\
A_{\mu}&2&-52&c_1&6&-c_0+c_1+c_3/2&15&-c_0+2c_1+c_3/2&21\\
A&1&4&-c_0+c_1+c_3/2&15&2c_0-2c_1+2c_2&20&c_0-c_1+2c_2+c_3/2&35\\
&&&&&&&\\
total~f&&&2(2c_0+2c_1+2c_2+c_3)&128&2(2c_0+2c_1+2c_2+c_3)&128&4(2c_0+2c_1+2c_2+c_3)&256\\
total~\tilde A&&&(65c_0-15c_1+c_2+c_3/2)/24&0&(-63c_0+13c_1+c_2-3c_3/2)/24&0&(2c_0-2c_1+2c_2-c_3)/24&0\\
\end{array}$
\caption{ ${\tilde X}^6$ compactification of $D=10$ Type IIA supergravity}
\label{D=10A'}
\end{table}
\subsection{{\cal N}=1}
The massless sectors of the $(\mathcal{N}=1,D=10)$ supergravity compactifications are given just by the NS sector
\begin{equation}
f=2(2c_0+2c_1+2c_2+c_3)
\end{equation}
and
\begin{equation}
A=-\frac{1}{24}(65c_0-15c_1+c_2+c_3/2).
\end{equation}
They have vanishing anomaly when
\begin{equation}
(c_0, c_1, c_2, c_3)= (1, 2{\cal N}-2, n, 30{\cal N}-95-n)
\end{equation}
and
\begin{equation}
f=4(26{\cal N}-97+3n)
\end{equation}
so the only possibility is:   
  \begin{itemize}
  \item{${\cal N}=4$, $n=15$, $f=128$}
\end{itemize}

Next consider an $({\cal N}=1, D=10)$ vector multiplet $(A_M, \chi)$ with $f=8+8$ as shown in Table \ref{D=10'}.
\begin{table}[h!]
$\begin{array}{llrrrrrrrrrr}
&Field &f&360A&X^6&X^4 \times T^2& T^6\\
&&&&&&&\\
\bigskip
&&&&\\
A_{M}&A_{\mu}&2&-52&c_0&d_0&1\\
~&{A}_{}&1&4&c_1&2 d_0+d_1&6\\
\chi~&\chi&2&7&c_0+c_1/2&2d_0+d_1/2&4\\
&&&&&\\
total~ f&&&&2(2c_0+c_1)&2(4d_0+d_1)&16&\\
&&&&&&&\\
total~A&&&&(-3c_0+c_1/2)/24&(-2d_0+d_1/2)/24&0\\
\end{array}$
\caption{ Compactifications of ${\cal N}$=1 $D=10$ vector multiplet}
\label{D=10'}
\end{table}

The massless sectors of the vector compactifications have 
\begin{equation}
f=2(2c_0+c_1)
\end{equation}
and
\begin{equation}
A=\frac{1}{24}(-3c_0+c_1/2).
\end{equation}
They have vanishing anomaly when
\begin{equation}
(c_0, c_1)= (1, 2{\cal N}-2)
\end{equation}
and
\begin{equation}
f=4{\cal N}.
\end{equation}
so the only possibility is:   
  \begin{itemize}
  \item{${\cal N}=4$, $f=16$}
\end{itemize}

\FloatBarrier
\subsection{ IIB }
\label{IIB}

Consider Type IIB in $D=10$. In the NS sector we have the  $D=10$ fields $(g_{MN}, \psi_M, B_{MN},\chi, \Phi)$; with $f=64+64$; in the R-R we have  $(A_{MNPQ}{}^+, \psi, C_{MN},\chi, D)$ also with $f=64+64$. If IIB is T-dual to IIA, we might expect $A(IIB)=\chi/24$ on $X^6$ since $A(IIA)=-\chi/24$ on $X^6$ and $\tilde A({ {IIA}})=\chi/24$ on its mirror.  But IIB is tricky: how do we assign four dimensional tensors coming from the self-dual 5-form in $D=10$? The authors of  \cite{Cremmer:1997ct,Cremmer:1998px}  address this problem in the case of $T^6$ by first writing the Lagrangian in $D=9$, where it coincides with that of IIA except  $A_{MNP}$ is swapped for its dual $A_{MNPQ}$, and then compactifying on $T^5$. The results are shown in Table \ref{D=10B} and, assigning  $360A=1080$ to $A_{\mu\nu\rho\sigma}$  as in section \ref {trace}, we find that the anomaly vanishes for IIB just as for IIA. Unfortunately, this trick does not generalize in a useful way for us, because $X^5 \times S^1$ has vanishing Euler number and is therefore not a good laboratory for testing mirror symmetry.

\begin{table}[h!]
$\begin{array}{llrrrrrrrrrrrrrrr}
Field &f&360A&NS&T^6&RR&T^6&IIB&T^6\\
&&&&&&&\\
\bigskip
&&&&\\
g_{\mu\nu}&2&848&&1&&0&&1\\
{\cal A}_{\mu}&2&-52&&6&&1&&7\\
{\cal A}&1&4&&21&&6&&27\\
\Phi&1&4&&1&&0&&1\\
\psi_{\mu}&2&-233&&4&&4&&8\\
\chi&2&7&&28&&28&&56\\
A_{\mu\nu\rho\sigma}&0&1080&&0&&1&&1\\
A_{\mu\nu\rho}&0&-720&&0&&5&&5\\
A_{\mu\nu}&1&364&&1&&11&&12\\
A_{\mu}&2&-52&&6&&15&&21\\
A&1&4&&15&&15&&30\\
&&&&&\\
total~f&&&&128&&128&&256\\
total~A&&&&0&&0&&0\\
\end{array}$
\caption{ $T^6$ compactification of $D=10$ Type IIB supergravity}
\label{D=10B}
\end{table}


\FloatBarrier
\section{Fermionic mirrors}
\label{Fermion}
The bosonic mirror map
\begin{equation}
(b_0,  b_1, b_2, b_3) \rightarrow (b_0, b_1,  b_2 -\rho/2, b_3+\rho/2)
\label{eq:mirrorbose}
\end{equation}
preserves the number of spin 3/2 and spin 1/2  but changes the number of spin 1 and spin 0 
as in  \autoref{boson}.  Note incidentally that the number of fields of spin $(2, 3/2, 1, 1/2, 0)$ equals $(b_0, b_0+b_1, b_1+b_2, b_2+b_3, 2b_3)$=$(a_0, a_1, a_2, a_3, a_4)$ where the $a_i$ are the betti numbers of $X^7 \times S^1$. Since there are equal numbers of bosons and fermions, we must have $2a_0-2a_1+2a_2-2a_3+a_4=0$ which is just the vanishing of the euler number $\chi(X^7 \times S^1)$.

\begin{table}[h!]
$\begin{array}{lllllllllrrrrrrr}
Spin &&&X&X'\\
&&&&\\
2&&&b_0&b_0\\
3/2&&&b_0+b_1&b_0+b_1\\
1&&&b_1+b_2&b_1+b_2 -\rho/2\\
1/2&&&b_2+b_3&b_2+b_3\\
0&&&2b_3&2b_3+\rho \\
&&\\
&&&f=4(b_0+b_1+b_2+b_3)&f'=4(b_0+b_1+b_2+b_3)&\\
&&&\rho=(7-5b_1+3b_2-b_3)&\rho'=-(7-5b_1+3b_2-b_3)&\\
\end{array}$
\caption{ Bosonic mirror symmetry}
\label{boson}
\end{table}

We define a fermionic mirror map
\begin{equation}
(b_0,  b_1, b_2, b_3) \rightarrow (b_0, b_1+{\cal N}'-{\cal N},  b_2 -{\cal N}'+{\cal N}, b_3)
\label{eq:mirrorfermi2}
\end{equation}
which preserves the number of spin 2, spin 1 and spin 0 but changes the number of spin 3/2 (from  ${\cal N} $ to ${\cal N}'$) and spin 1/2, keeping $f$ fixed as in  \autoref{fermion}.  
 \begin{table}[h!]
$\begin{array}{lllllllrrrrrrr}
Spin &&&X&X'\\
&&&&\\
2&&&b_0&b_0\\
3/2&&&b_0+b_1&b_0+b_1+{\cal N}'-{\cal N}\\
1&&&b_1+b_2&b_1+b_2\\
1/2&&&b_2+b_3&b_2+b_3-{\cal N}'+{\cal N}\\
0&&&2b_3&2b_3 \\
&&\\
&&&f=4(1+b_1+b_2+b_3)&f'=4(1+b_1+b_2+b_3)&\\
&&&\rho=(7-5b_1+3b_2-b_3)&\rho'=(7-5b_1+3b_2-b_3)-8(N'-{\cal N})&\\
\end{array}$
\caption{ Fermionic mirror symmtry}
\label{fermion}
\end{table}

However, we further require that each member of the pair have identical bosonic lagrangians. Previously known examples \cite{Andrianopoli:1996wf,Ferrara:2008ap,Roest:2009sn} are provided by $({\cal N}=6,{\cal N}'=2)$, $({\cal N}=4,{\cal N}'=2)$ and  $({\cal N}=3,{\cal N}'=2)$ supergravity plus matter theories, as shown below. We also provide the relevant coset structure (after dualization).  Curiously, the partner with the higher supersymmetry is generalized self-mirror in the bosonic sense. In addition,  we  find a new two-parameter family with $({\cal N}=1,{\cal N}'=2)$.
\begin{itemize}
\item{${\cal N}=6$, ${\cal N}'$=2}

${\cal N}=6$  with $(b_0,b_1,b_2,b_3)=(1,5,11,15)$ and Magic ${\cal N}'$=2 with $(b_0,b_1,b_2,b_3)=(1,1,15,15)$ as in \autoref{magic6}.
\begin{table}[h!]
$\begin{array}{lllllllrrrrrrr}
Spin &&&X&X'\\
&&&&\\
2&&&1&1\\
3/2&&&6&2\\
1&&&16&16\\
1/2&&&26&30\\
0&&&30&30 \\
&&&&\\
&&&f=128&f'=128\\
&&&\rho=0&\rho'=32\\
\end{array}$
\caption{$ {\cal N}=6$ and $ {\cal N}'=2$}
\label{magic6}
\end{table}

The relevant coset is
\be
\frac{SO^*(12)}{U(6)}
\ee

\item{${\cal N}=4$, ${\cal N'}$=2}

${\cal N}=4$  with $(b_0,b_1,b_2,b_3)=(1,3,5,7)$ and ${\cal N'}$=2 with $(b_0,b_1,b_2,b_3)=(1,1,7,7)$ as in \autoref{4}.
\begin{table}[h!]
$\begin{array}{lllllllrrrrrrr}
Spin &&&X&X'\\
&&&&\\
2&&&1&1\\
3/2&&&4&2\\
1&&&8&8\\
1/2&&&12&14\\
0&&&14&14 \\
&&&&\\
&&&f=64&f'=64\\
&&&\rho=0&\rho'=16\\
\end{array}$
\caption{$ {\cal N}=4$ and ${\cal N}'=2$}
\label{4}
\end{table}

The relevant coset is
\be
\frac{SL(2)}{U(1)} \times \frac{SO(6,2)}{SO(6) \times SO(2)}
\ee

\item{${\cal N}=3$, ${\cal N}'$=2}

${\cal N}=3$  with $(b_0,b_1,b_2,b_3)=(1,2,2,3)$ and ${\cal N}'$=2 with $(b_0,b_1,b_2,b_3)=(1,1,3,3)$ as in \autoref{3}.
\begin{table}[h!]
$\begin{array}{lllllllrrrrrrr}
Spin &&&X&X'\\
&&&&\\
2&&&1&1\\
3/2&&&3&2\\
1&&&4&4\\
1/2&&&5&6\\
0&&&6&6\\
&&&&\\
&&&f=32&f'=32\\
&&&\rho=0&\rho'=8\\
\end{array}$
\caption{ ${\cal N}=4$ and ${\cal N}'=2$}
\label{3}
\end{table}

The relevant coset is
\be
\frac{SU(3,1)}{SU(3) \times U(1)} 
\ee

\item{${\cal N}=1$, ${\cal N}'$=2}

${\cal N}=1$  with $(b_0,b_1,b_2,b_3)=(1,0,n_1+1,2n_2+n_1)$ and ${\cal N}'$=2 with $(b_0,b_1,b_2,b_3)=(1,1,n_1, 2n_2+n_1)$ as in \autoref{1}.
\begin{table}[h!]
$\begin{array}{lllllllrrrrrrr}
Spin &&&X&X'\\
&&&&\\
2&&&1&1\\
3/2&&&1&2\\
1&&&n_1+1&n_1+1\\
1/2&&&2n_2+2n_1+1&2n_2+2n_1\\
0&&&4n_2+2n_1&4n_2+2n_1 \\
&&&&\\
&&&f=8(n_1+n_2+1)&f'=8(n_1+n_2+1)\\
&&&\rho=2(1+n_1-n_2)+8&\rho'=2(1+n_1-n_2)\
\end{array}$
\caption{ ${\cal N}=1$ and ${\cal N}'=2$}
\label{1}
\end{table}

The relevant coset is
\be
\frac{SU(1,n_1)}{U(n_1)} \times \frac{SU(2,n_2)}{SU(2) \times SU(n_2) \times U(1)} 
\ee
and describes $\mathcal{N}=1$ supergravity plus $n_1+1$ vector mutiplets and $n_1+2n_2$ chiral paired with $\mathcal{N}=2$ supergravity plus $n_1$ vector multiplets and $n_2$ hypermutiplets. The $\mathcal{N}=1$ partner is bosonic self mirror when $n_2=n_1+5$ and the $\mathcal{N}=2$ partner is bosonic is self-mirror when $n_2=n_1+1$.  

\end {itemize}

\section{Acknowledgments}

Conversations and correspondence with Leron Borsten, Duminda Dahanayke, Alessio Marrani, William Rubens,  Ashoke Sen, Samson Shatashvili and Edward Witten are much appreciated. MJD is supported in part by the STFC under rolling Grant No. ST/G000743/1.  S. F. is supported by the ERC Advanced Grant no 226455,``Supersymmetry, Quantum Gravity and Gauge Fields" and in part by DOE Grant DE-FG03-91ER40662.  MJD is grateful for hospitality at the CERN theory division, where he was supported by the above ERC Advanced Grant.

 \FloatBarrier
 \appendix

\FloatBarrier
\section{Dualization}
\label{dual}
\subsection{$\mathcal{N}=1$,$D=11$}
In this section we list what the anomalies would have been had we used the dualized form of the theories where $A_{\mu\nu\rho}$ is 
set to zero and $A_{\mu\nu}$ is replaced by ${\cal A}$. The results, shown in Table \ref{D=11'} agree with those in 
\cite{Christensen:1980ee} and are also what would be obtained for the undualized version if one took the view that the anomaly for $A_{\mu\nu}$ is equal to that of ${\cal A}$ and the anomaly for $A_{\mu\nu\rho}$ is zero \cite{Siegel:1980ax,Banerjee:2010qc}.  In Tables \ref{Y7'} to \ref{Y0'},  we group the individual fields into supermultiplets after dualization.

\begin{table}[h!]
$\begin{array}{lrrrrrrrrrrrrrr}
Spin &f&360A&X^7&X^6 \times S^1&X^5 \times T^2&X^4 \times T^3 &X^3 \times T^4&X^2 \times T^5&T^7 \\
&&&&\\
2&2&848&b_0&c_0&e_0&d_0&j_0&k_0&1\\
3/2&2&-233&b_0+b_1&2c_0+c_1&3e_0+e_1&4d_0+d_1&5j_0+j_1&6k_0+k_1&8\\
1&2&-52&b_1+b_2&c_0+2c_1+c_2&3e_0+3e_1+e_2&6d_0+4d_1+d_2&5j_0+j_1&16k_0+6k_1&28\\
1/2&2&7&b_2+b_3&c_1+2c_2+c_3&e_0+3e_1+4e_2&4d_0+7d_1+4d_2&11j_0+5j_1&26k_0+15k_1&56\\
0&1&4&2b_3&2c_2+2c_3&2e_1+6e_2&2d_0+8d_1+6d_2&10j_0+20j_1&30k_0+20k_1&70\\
&&\\
\end{array}$
\caption{ $X^{(8-{\cal N})}\times T^{({\cal N}-1)}$ compactification of $D=11$ supergravity after dualization}\label{D=11'}
\end{table}

\begin{table}[h!]
$\begin{array}{llrrrrrrrrrr}
{\cal N}=1~&multiplet  &f&360A&{\cal N}=b_0+b_1 &  {\cal N}=8 &  {\cal N}=1 & \\
&&&&&&&\\
\bigskip
graviton&(g_{\mu\nu};\psi_{\mu})&2+2&615&b_0&1&1\\
gravitino&({\cal A}_{\mu}; \psi_{\mu}) &2+2&-285&b_1&7&0 \\
&&&&&&&\\
vector&(\chi; A_{\mu})&2+2&-45&b_2&21&b_2\\
&&&&&&&\\
 chiral &({\cal A}; \chi; A)&2+2&15&b_3&35&b_3\\
&&&&&&\\
&&&&&&\\
total ~f &&&&4(b_0+b_1+b_2+b_3)&256&4(1+b_2+b_3)\\
&&&&&&\\
total ~A &&&&(41b_0-19b_1-3b_2+b_3)/24&-5&(41-3b_2+b_3)/24\\
\end{array}$
\caption{ The $D=4$ multiplets in an  ${\cal N}$=1 basis after dualization. }
\label{Y7'}
\end{table}

\begin{table}[h!]
$\begin{array}{llrrrrrrrrrrrrr}
{\cal N}=2~& multiplet  &f&360A&{\cal N}=2c_0+c_1 &  {\cal N}=8 &  {\cal N}=2 & \\
&&&&&&&\\
\bigskip
graviton&(g_{\mu\nu}, {\cal A}_{\mu}; 2\psi_{\mu})&4+4&330&c_0&1&1\\
gravitino&({\cal A}_{\mu}; \psi_{\mu}, \chi; A_{\mu}) &4+4&-330&c_1&6&0 \\
&&&&&&&\\
vector&({\cal A}, 2\chi; A_{\mu}, A)&4+4&-30&c_2&15&c_2\\
&&&&&&&\\
hyper&(2{\cal A}; 2\chi; 2A)&4+4&30&c_3/2&10&c_3/2\\
&&&&&&\\
&&&&&&\\
total ~f &&&&4(2c_0+2c_1+2c_2+c_3)&256&4(2+2c_2+c_3)\\
&&&&&&\\
total ~A &&&&(22c_0-22c_1-2c_2+c_3)/24&-5&(22-2c_2+c_3)/24\\
\end{array}$
\caption{  The $D=4$ multiplets in an  ${\cal N}$=2 basis after dualization.}
\label{Y6'}
\end{table}

\begin{table}[h!]
$\begin{array}{llrrrrrrrrrr}
{\cal N}=3~& multiplet  &f&360A&{\cal N}=3e_0+e_1 &  {\cal N}=8 & {\cal N}=3 & \\
&&&&&&&\\
\bigskip
graviton&(g_{\mu\nu}, 2{\cal A}_{\mu};  3\psi_{\mu}, \chi;  A_{\mu})&8+8&0&e_0&1&1\\
gravitino&({\cal A}_{\mu}; \psi_{\mu}, {\cal A}: 3\chi;  2A_{\mu},A) &8+8&-360&e_1&5&0 \\
&&&&&&&\\
vector&(3{\cal A}; 4\chi; A_{\mu}, 3A)&8+8&0&e_2&10&e_2\\
&&&&&&\\
&&&&&&&\\
total ~f &&&&16(e_0+e_1+e_2)&256&16(1+e_2)\\
&&&&&&\\
total ~A &&&&-e_1&-5&0\\
\end{array}$
\caption{   The $D=4$ multiplets in an  ${\cal N}$=3 basis after dualization.}
\label{Y5'}
\end{table}

\begin{table}[h!]
$\begin{array}{llrrrrrrrrrr}
{\cal N}=4~  & multiplet&f&360A& {\cal N}=4d_0+d_1 &  {\cal N}=8 &  {\cal N}=4 & \\
&&&&&&&\\
\bigskip
graviton&(g_{\mu\nu}, 3 {\cal A}_{\mu},  {\cal A}, 4 \psi_{\mu}, 4 \chi, 3  A_{\mu}, A)&16+16&-360&d_0&1&1\\
gravitino&({\cal A}_{\mu},  4{\cal A},  \psi_{\mu}, 7 \chi, 3A_{\mu},4A) &16+16&-360&d_1&4&0 \\
&&&&&&&\\
vector&(3{\cal A};  4 \chi; A_{\mu}, 3 A)&8+8&0&d_2&6&d_2\\
&&&&&\\
&&&&&&&\\
total ~f &&&&16(2d_0+2d_1+d_2)&256&16(1+d_2)\\
&&&&&&\\
total ~A &&&&-(d_0+d_1)&-5&-1\\
\end{array}$
\caption{  The $D=4$ multiplets in an  ${\cal N}$=4 basis after dualization.}
\label{Y4'}
\end{table}

\begin{table}[h!]
$\begin{array}{llrrrrrrrrrr}
{\cal N}=5~  & multiplet&f&360A&{\cal N}=5j_0+j_1&{\cal N}=8&{\cal N}=5  \\
&&&&&&&\\
\bigskip
graviton&(g_{\mu\nu}, 4{\cal A}_{\mu}, 5{\cal A}; 5\psi_{\mu}, 11\chi: 6A_{\mu}, 5A)
&32+32&-720&j_0&1&1\\
gravitino&({\cal A}_{\mu},  10{\cal A},  \psi_{\mu}, 15 \chi,  5A_{\mu}, 10A) &32+32&-360&j_1&3&0 \\
&&&&&&&\\
total ~f &&&&64(j_0+j_1)&256&64\\
&&&&&&&\\
total ~A &&&&-2j_0-j_1&-5&-2\\
&&&&&&\\
\end{array}$
\caption{ The $D=4$ multiplets in an  ${\cal N}$=5 basis after dualization.}
\label{Y3'}
\end{table}

\begin{table}[h!]
$\begin{array}{llrrrrrrrrrr}
{\cal N}=6~  & multiplet&f&360A&{\cal N}=6+k_1&{\cal N}=8&{\cal N}=6  \\
&&&&&&&\\
\bigskip
graviton&(g_{\mu\nu}, 5{\cal A}_{\mu}, 15{\cal A}; 6\psi_{\mu}, 26\chi; 11A_{\mu},15A)
&64+64&-1080&k_0&1&1\\
gravitino&({\cal A}_{\mu},  10{\cal A},  \psi_{\mu}, 15\chi; 5A_{\mu},10A) &32+32&-360&k_1&2&0 \\
&&&&&&&\\
total ~f &&&&64(2k_0+k_1)&256&128\\
&&&&&&&\\
total ~A &&&&-3k_0-k_1&-5&-3\\
\end{array}$
\caption{  The $D=4$ multiplets in an  ${\cal N}$=6 basis after dualization.}
\label{Y2'}
\end{table}

\begin{table}[h!]
$\begin{array}{llrrrrrrrrrr}
{\cal N}=8~  & multiplet&f&360A&{\cal N}=8  \\
&&&&&&&\\
\bigskip
graviton&(g_{\mu\nu}, 7{\cal A}_{\mu}, 35 {\cal A}; 8\psi_{\mu}, 56\chi; 21 A_{\mu},35 A)
&256&-1800&1&\\
&&&&&&&\\  
total ~f &&&&256\\
&&&&&&&\\
total ~A &&&&-5\\
\end{array}$
\caption{ The $D=4$ multiplets in an  ${\cal N}$=8 basis after dualization.}
\label{Y0'}
\end{table}

\FloatBarrier
\subsection{Type IIA  and IIB on $X^6$}

We have 
\begin{equation}
A(IIA)-A(IIB)=(-4c_0+4c_1-4c_2+2c_3)/24=-\frac{\chi}{12}
\end{equation}

\begin{table}[h!]
$\begin{array}{lrrrrrrrrrrrrrr}
Spin &f&360A&NS&T^6&RR&T^6&IIA&T^6 \\
&&&&\\
2&2&848&c_0&1&0&0&c_0&1\\
3/2&2&-233&c_0+c_1/2&4&c_0+c_1/2&4&2c_0+c_1&8\\
1&2&-52&2c_1&12&c_0+c_2&16&c_0+2c_1+c_2&28\\
1/2&2&7&c_1/2+c_2+c_3/2&28&c_1/2+c_2+c_3/2&28&c_1+2c_2+c_3&56\\
0&1&4&-2c_1+2c_2+c_3&38&2c_1+c_3&32&2c_2+2c_3&70\\
&&\\
total~f&&&2(2c_0+2c_1+2c_2+c_3)&128&2(2c_0+2c_1+2c_2+c_3)&128&4(2c_0+2c_1+2c_2+c_3)&256\\
&&&&&&&\\
total~A&&&(41c_0-15c_1+c_2+c_3/2)/24&-1&(-19c_0-7c_1-3c_2+c_3/2)/24&-4&(22c_0-22c_1-2c_2+c_3)/24&-5\\
\end{array}$
\caption{ $X^6$ compactification of $D=10$ Type IIA supergravity after dualization}
\label{D=10Aafter}
\end{table}

\begin{table}[h!]
$\begin{array}{lrrrrrrrrrrrrrr}
Spin &f&360A&NS&T^6&RR&T^6&IIB&T^6\\
&&&&\\
2&2&848&c_0&1&0&0&c_0&1\\
3/2&2&-233&c_0+c_1/2&4&c_0+c_1/2&4&2c_0+c_1&8\\
1&2&-52&2c_1&12&c_1+c_3/2&16&3c_1+c_3/2&28\\
1/2&2&7&c_1/2+c_2+c_3/2&28&c_1/2+c_2+c_3/2&28&c_1+2c_2+c_3&56\\
0&1&4&-2c_1+2c_2+c_3&38&2c_0+2c_2&32&2c_0-2c_1+4c_2+c_3&70\\
&&\\
total~f&&&2(2c_0+2c_1+2c_2+c_3)&128&2(2c_0+2c_1+2c_2+c_3)&128&4(2c_0+2c_1+2c_2+c_3)&256\\
&&&&&&&\\
total~A&&&(41c_0-15c_1+c_2+c_3/2)/24&-1&(-15c_0-11c_1+c_2-3c_3/2)/24&-4&(26c_0-26c_1+2c_2-c_3)/24&-5\\
\end{array}$
\caption{ $X^6$ compactification of $D=10$ Type IIB supergravity after dualization}
\label{D=10Bafter}
\end{table}
\FloatBarrier
\bibliographystyle{apsrev}
\bibliography{arxiv,notarxiv}

\begin{thebibliography}{34}
\expandafter\ifx\csname natexlab\endcsname\relax\def\natexlab#1{#1}\fi
\expandafter\ifx\csname bibnamefont\endcsname\relax
  \def\bibnamefont#1{#1}\fi
\expandafter\ifx\csname bibfnamefont\endcsname\relax
  \def\bibfnamefont#1{#1}\fi
\expandafter\ifx\csname citenamefont\endcsname\relax
  \def\citenamefont#1{#1}\fi
\expandafter\ifx\csname url\endcsname\relax
  \def\url#1{\texttt{#1}}\fi
\expandafter\ifx\csname urlprefix\endcsname\relax\def\urlprefix{URL }\fi
\providecommand{\bibinfo}[2]{#2}
\providecommand{\eprint}[2][]{\url{#2}}

\bibitem[{\citenamefont{Joyce}(1996{\natexlab{a}})}]{Joyce:1996a}
\bibinfo{author}{\bibfnamefont{D.}~\bibnamefont{Joyce}}, \bibinfo{journal}{J.
  Differential Geometry} \textbf{\bibinfo{volume}{43}}
  (\bibinfo{year}{1996}{\natexlab{a}}).

\bibitem[{\citenamefont{Joyce}(1996{\natexlab{b}})}]{Joyce:1996b}
\bibinfo{author}{\bibfnamefont{D.}~\bibnamefont{Joyce}}, \bibinfo{journal}{J.
  Differential Geometry} \textbf{\bibinfo{volume}{43}}, \bibinfo{pages}{329}
  (\bibinfo{year}{1996}{\natexlab{b}}).

\bibitem[{\citenamefont{Shatashvili and Vafa}(1995)}]{Shatashvili:1994zw}
\bibinfo{author}{\bibfnamefont{S.~L.} \bibnamefont{Shatashvili}}
  \bibnamefont{and} \bibinfo{author}{\bibfnamefont{C.}~\bibnamefont{Vafa}},
  \bibinfo{journal}{Selecta Math.} \textbf{\bibinfo{volume}{1}},
  \bibinfo{pages}{347} (\bibinfo{year}{1995}), \eprint{hep-th/9407025}.

\bibitem[{\citenamefont{Acharya}(1998)}]{Acharya:1997rh}
\bibinfo{author}{\bibfnamefont{B.~S.} \bibnamefont{Acharya}},
  \bibinfo{journal}{Nucl. Phys.} \textbf{\bibinfo{volume}{B524}},
  \bibinfo{pages}{269} (\bibinfo{year}{1998}), \eprint{hep-th/9707186}.

\bibitem[{\citenamefont{Gaberdiel and Kaste}(2004)}]{Gaberdiel:2004vx}
\bibinfo{author}{\bibfnamefont{M.~R.} \bibnamefont{Gaberdiel}}
  \bibnamefont{and} \bibinfo{author}{\bibfnamefont{P.}~\bibnamefont{Kaste}},
  \bibinfo{journal}{JHEP} \textbf{\bibinfo{volume}{08}}, \bibinfo{pages}{001}
  (\bibinfo{year}{2004}), \eprint{hep-th/0401125}.

\bibitem[{\citenamefont{Duff}(1977)}]{Duff:1977ay}
\bibinfo{author}{\bibfnamefont{M.~J.} \bibnamefont{Duff}},
  \bibinfo{journal}{Nucl. Phys.} \textbf{\bibinfo{volume}{B125}},
  \bibinfo{pages}{334} (\bibinfo{year}{1977}).

\bibitem[{\citenamefont{Duff}(1994)}]{Duff:1993wm}
\bibinfo{author}{\bibfnamefont{M.~J.} \bibnamefont{Duff}},
  \bibinfo{journal}{Class. Quant. Grav.} \textbf{\bibinfo{volume}{11}},
  \bibinfo{pages}{1387} (\bibinfo{year}{1994}), \eprint{hep-th/9308075}.

\bibitem[{\citenamefont{Duff et~al.}(1983)\citenamefont{Duff, Nilsson, and
  Pope}}]{Duff:1983vj}
\bibinfo{author}{\bibfnamefont{M.~J.} \bibnamefont{Duff}},
  \bibinfo{author}{\bibfnamefont{B.~E.~W.} \bibnamefont{Nilsson}},
  \bibnamefont{and} \bibinfo{author}{\bibfnamefont{C.~N.} \bibnamefont{Pope}},
  \bibinfo{journal}{Phys. Lett.} \textbf{\bibinfo{volume}{B129}},
  \bibinfo{pages}{39} (\bibinfo{year}{1983}).

\bibitem[{\citenamefont{Duff and Ferrara}(2010)}]{Duff:2010b}
\bibinfo{author}{\bibfnamefont{M.~J.} \bibnamefont{Duff}} \bibnamefont{and}
  \bibinfo{author}{\bibfnamefont{S.}~\bibnamefont{Ferrara}}
  (\bibinfo{year}{2010}).

\bibitem[{\citenamefont{Cremmer and Julia}(1979)}]{Cremmer:1979up}
\bibinfo{author}{\bibfnamefont{E.}~\bibnamefont{Cremmer}} \bibnamefont{and}
  \bibinfo{author}{\bibfnamefont{B.}~\bibnamefont{Julia}},
  \bibinfo{journal}{Nucl. Phys.} \textbf{\bibinfo{volume}{B159}},
  \bibinfo{pages}{141} (\bibinfo{year}{1979}).

\bibitem[{\citenamefont{Ferrara and Kounnas}(1989)}]{Ferrara:1989nm}
\bibinfo{author}{\bibfnamefont{S.}~\bibnamefont{Ferrara}} \bibnamefont{and}
  \bibinfo{author}{\bibfnamefont{C.}~\bibnamefont{Kounnas}},
  \bibinfo{journal}{Nucl. Phys.} \textbf{\bibinfo{volume}{B328}},
  \bibinfo{pages}{406} (\bibinfo{year}{1989}).

\bibitem[{\citenamefont{Sen and Vafa}(1995)}]{Sen:1995ff}
\bibinfo{author}{\bibfnamefont{A.}~\bibnamefont{Sen}} \bibnamefont{and}
  \bibinfo{author}{\bibfnamefont{C.}~\bibnamefont{Vafa}},
  \bibinfo{journal}{Nucl. Phys.} \textbf{\bibinfo{volume}{B455}},
  \bibinfo{pages}{165} (\bibinfo{year}{1995}), \eprint{hep-th/9508064}.

\bibitem[{\citenamefont{Tung}(1992)}]{Shing:492595}
\bibinfo{author}{\bibfnamefont{Y.~S.} \bibnamefont{Tung}},
  \emph{\bibinfo{title}{Essays on mirror manifolds}} (\bibinfo{publisher}{Int.
  Press}, \bibinfo{address}{Hong Kong}, \bibinfo{year}{1992}).

\bibitem[{\citenamefont{Duff and van Nieuwenhuizen}(1980)}]{Duff:1980qv}
\bibinfo{author}{\bibfnamefont{M.~J.} \bibnamefont{Duff}} \bibnamefont{and}
  \bibinfo{author}{\bibfnamefont{P.}~\bibnamefont{van Nieuwenhuizen}},
  \bibinfo{journal}{Phys. Lett.} \textbf{\bibinfo{volume}{B94}},
  \bibinfo{pages}{179} (\bibinfo{year}{1980}).

\bibitem[{\citenamefont{Witten}(1995)}]{Witten:1995gf}
\bibinfo{author}{\bibfnamefont{E.}~\bibnamefont{Witten}},
  \bibinfo{journal}{Selecta Math.} \textbf{\bibinfo{volume}{1}},
  \bibinfo{pages}{383} (\bibinfo{year}{1995}), \eprint{hep-th/9505186}.

\bibitem[{\citenamefont{Nicolai and Townsend}(1981)}]{Nicolai:1980td}
\bibinfo{author}{\bibfnamefont{H.}~\bibnamefont{Nicolai}} \bibnamefont{and}
  \bibinfo{author}{\bibfnamefont{P.~K.} \bibnamefont{Townsend}},
  \bibinfo{journal}{Phys. Lett.} \textbf{\bibinfo{volume}{B98}},
  \bibinfo{pages}{257} (\bibinfo{year}{1981}).

\bibitem[{\citenamefont{Siegel}(1981)}]{Siegel:1980ax}
\bibinfo{author}{\bibfnamefont{W.}~\bibnamefont{Siegel}},
  \bibinfo{journal}{Phys. Lett.} \textbf{\bibinfo{volume}{B103}},
  \bibinfo{pages}{107} (\bibinfo{year}{1981}).

\bibitem[{\citenamefont{Fradkin and Tseytlin}(1984)}]{Fradkin:1983tg}
\bibinfo{author}{\bibfnamefont{E.~S.} \bibnamefont{Fradkin}} \bibnamefont{and}
  \bibinfo{author}{\bibfnamefont{A.~A.} \bibnamefont{Tseytlin}},
  \bibinfo{journal}{Phys. Lett.} \textbf{\bibinfo{volume}{B134}},
  \bibinfo{pages}{187} (\bibinfo{year}{1984}).

\bibitem[{\citenamefont{Grisaru et~al.}(1984)\citenamefont{Grisaru, Nielsen,
  Siegel, and Zanon}}]{Grisaru:1984vk}
\bibinfo{author}{\bibfnamefont{M.~T.} \bibnamefont{Grisaru}},
  \bibinfo{author}{\bibfnamefont{N.~K.} \bibnamefont{Nielsen}},
  \bibinfo{author}{\bibfnamefont{W.}~\bibnamefont{Siegel}}, \bibnamefont{and}
  \bibinfo{author}{\bibfnamefont{D.}~\bibnamefont{Zanon}},
  \bibinfo{journal}{Nucl. Phys.} \textbf{\bibinfo{volume}{B247}},
  \bibinfo{pages}{157} (\bibinfo{year}{1984}).

\bibitem[{\citenamefont{Buchbinder et~al.}(2008)\citenamefont{Buchbinder,
  Kirillova, and Pletnev}}]{Buchbinder:2008jf}
\bibinfo{author}{\bibfnamefont{I.~L.} \bibnamefont{Buchbinder}},
  \bibinfo{author}{\bibfnamefont{E.~N.} \bibnamefont{Kirillova}},
  \bibnamefont{and} \bibinfo{author}{\bibfnamefont{N.~G.}
  \bibnamefont{Pletnev}}, \bibinfo{journal}{Phys. Rev.}
  \textbf{\bibinfo{volume}{D78}}, \bibinfo{pages}{084024}
  (\bibinfo{year}{2008}), \eprint{0806.3505}.

\bibitem[{\citenamefont{Antoniadis et~al.}(1992)\citenamefont{Antoniadis, Gava,
  and Narain}}]{Antoniadis:1992sa}
\bibinfo{author}{\bibfnamefont{I.}~\bibnamefont{Antoniadis}},
  \bibinfo{author}{\bibfnamefont{E.}~\bibnamefont{Gava}}, \bibnamefont{and}
  \bibinfo{author}{\bibfnamefont{K.~S.} \bibnamefont{Narain}},
  \bibinfo{journal}{Phys. Lett.} \textbf{\bibinfo{volume}{B283}},
  \bibinfo{pages}{209} (\bibinfo{year}{1992}), \eprint{hep-th/9203071}.

\bibitem[{\citenamefont{Fursaev}(1995)}]{Fursaev:1994te}
\bibinfo{author}{\bibfnamefont{D.~V.} \bibnamefont{Fursaev}},
  \bibinfo{journal}{Phys. Rev.} \textbf{\bibinfo{volume}{D51}},
  \bibinfo{pages}{5352} (\bibinfo{year}{1995}), \eprint{hep-th/9412161}.

\bibitem[{\citenamefont{Solodukhin}(2008)}]{Solodukhin:2008dh}
\bibinfo{author}{\bibfnamefont{S.~N.} \bibnamefont{Solodukhin}},
  \bibinfo{journal}{Phys. Lett.} \textbf{\bibinfo{volume}{B665}},
  \bibinfo{pages}{305} (\bibinfo{year}{2008}), \eprint{0802.3117}.

\bibitem[{\citenamefont{Banerjee et~al.}(2010)\citenamefont{Banerjee, Gupta,
  and Sen}}]{Banerjee:2010qc}
\bibinfo{author}{\bibfnamefont{S.}~\bibnamefont{Banerjee}},
  \bibinfo{author}{\bibfnamefont{R.~K.} \bibnamefont{Gupta}}, \bibnamefont{and}
  \bibinfo{author}{\bibfnamefont{A.}~\bibnamefont{Sen}} (\bibinfo{year}{2010}),
  \eprint{1005.3044}.

\bibitem[{\citenamefont{Aros et~al.}(2010)\citenamefont{Aros, Diaz, and
  Montecinos}}]{Aros:2010jb}
\bibinfo{author}{\bibfnamefont{R.}~\bibnamefont{Aros}},
  \bibinfo{author}{\bibfnamefont{D.~E.} \bibnamefont{Diaz}}, \bibnamefont{and}
  \bibinfo{author}{\bibfnamefont{A.}~\bibnamefont{Montecinos}},
  \bibinfo{journal}{JHEP} \textbf{\bibinfo{volume}{07}}, \bibinfo{pages}{012}
  (\bibinfo{year}{2010}), \eprint{1003.1083}.

\bibitem[{\citenamefont{Dowker}(2010)}]{Dowker:2010bu}
\bibinfo{author}{\bibfnamefont{J.~S.} \bibnamefont{Dowker}}
  (\bibinfo{year}{2010}), \eprint{1009.3854}.

\bibitem[{\citenamefont{Ferrara et~al.}(2006)\citenamefont{Ferrara, Gimon, and
  Kallosh}}]{Ferrara:2006yb}
\bibinfo{author}{\bibfnamefont{S.}~\bibnamefont{Ferrara}},
  \bibinfo{author}{\bibfnamefont{E.~G.} \bibnamefont{Gimon}}, \bibnamefont{and}
  \bibinfo{author}{\bibfnamefont{R.}~\bibnamefont{Kallosh}},
  \bibinfo{journal}{Phys. Rev.} \textbf{\bibinfo{volume}{D74}},
  \bibinfo{pages}{125018} (\bibinfo{year}{2006}), \eprint{hep-th/0606211}.

\bibitem[{\citenamefont{Ferrara et~al.}(2008)\citenamefont{Ferrara, Gnecchi,
  and Marrani}}]{Ferrara:2008ap}
\bibinfo{author}{\bibfnamefont{S.}~\bibnamefont{Ferrara}},
  \bibinfo{author}{\bibfnamefont{A.}~\bibnamefont{Gnecchi}}, \bibnamefont{and}
  \bibinfo{author}{\bibfnamefont{A.}~\bibnamefont{Marrani}},
  \bibinfo{journal}{Phys. Rev.} \textbf{\bibinfo{volume}{D78}},
  \bibinfo{pages}{065003} (\bibinfo{year}{2008}), \eprint{0806.3196}.

\bibitem[{\citenamefont{Roest and Samtleben}(2009)}]{Roest:2009sn}
\bibinfo{author}{\bibfnamefont{D.}~\bibnamefont{Roest}} \bibnamefont{and}
  \bibinfo{author}{\bibfnamefont{H.}~\bibnamefont{Samtleben}},
  \bibinfo{journal}{Class. Quant. Grav.} \textbf{\bibinfo{volume}{26}},
  \bibinfo{pages}{155001} (\bibinfo{year}{2009}), \eprint{0904.1344}.

\bibitem[{\citenamefont{Siegel}(1980)}]{Siegel:1980jj}
\bibinfo{author}{\bibfnamefont{W.}~\bibnamefont{Siegel}},
  \bibinfo{journal}{Phys. Lett.} \textbf{\bibinfo{volume}{B93}},
  \bibinfo{pages}{170} (\bibinfo{year}{1980}).

\bibitem[{\citenamefont{Cremmer
  et~al.}(1998{\natexlab{a}})\citenamefont{Cremmer, Julia, Lu, and
  Pope}}]{Cremmer:1997ct}
\bibinfo{author}{\bibfnamefont{E.}~\bibnamefont{Cremmer}},
  \bibinfo{author}{\bibfnamefont{B.}~\bibnamefont{Julia}},
  \bibinfo{author}{\bibfnamefont{H.}~\bibnamefont{Lu}}, \bibnamefont{and}
  \bibinfo{author}{\bibfnamefont{C.~N.} \bibnamefont{Pope}},
  \bibinfo{journal}{Nucl. Phys.} \textbf{\bibinfo{volume}{B523}},
  \bibinfo{pages}{73} (\bibinfo{year}{1998}{\natexlab{a}}),
  \eprint{hep-th/9710119}.

\bibitem[{\citenamefont{Cremmer
  et~al.}(1998{\natexlab{b}})\citenamefont{Cremmer, Julia, Lu, and
  Pope}}]{Cremmer:1998px}
\bibinfo{author}{\bibfnamefont{E.}~\bibnamefont{Cremmer}},
  \bibinfo{author}{\bibfnamefont{B.}~\bibnamefont{Julia}},
  \bibinfo{author}{\bibfnamefont{H.}~\bibnamefont{Lu}}, \bibnamefont{and}
  \bibinfo{author}{\bibfnamefont{C.~N.} \bibnamefont{Pope}},
  \bibinfo{journal}{Nucl. Phys.} \textbf{\bibinfo{volume}{B535}},
  \bibinfo{pages}{242} (\bibinfo{year}{1998}{\natexlab{b}}),
  \eprint{hep-th/9806106}.

\bibitem[{\citenamefont{Andrianopoli et~al.}(1997)\citenamefont{Andrianopoli,
  D'Auria, and Ferrara}}]{Andrianopoli:1996wf}
\bibinfo{author}{\bibfnamefont{L.}~\bibnamefont{Andrianopoli}},
  \bibinfo{author}{\bibfnamefont{R.}~\bibnamefont{D'Auria}}, \bibnamefont{and}
  \bibinfo{author}{\bibfnamefont{S.}~\bibnamefont{Ferrara}},
  \bibinfo{journal}{Int. J. Mod. Phys.} \textbf{\bibinfo{volume}{A12}},
  \bibinfo{pages}{3759} (\bibinfo{year}{1997}), \eprint{hep-th/9608015}.

\bibitem[{\citenamefont{Christensen et~al.}(1980)\citenamefont{Christensen,
  Duff, Gibbons, and Rocek}}]{Christensen:1980ee}
\bibinfo{author}{\bibfnamefont{S.~M.} \bibnamefont{Christensen}},
  \bibinfo{author}{\bibfnamefont{M.~J.} \bibnamefont{Duff}},
  \bibinfo{author}{\bibfnamefont{G.~W.} \bibnamefont{Gibbons}},
  \bibnamefont{and} \bibinfo{author}{\bibfnamefont{M.}~\bibnamefont{Rocek}},
  \bibinfo{journal}{Phys. Rev. Lett.} \textbf{\bibinfo{volume}{45}},
  \bibinfo{pages}{161} (\bibinfo{year}{1980}).

\end{thebibliography}

\end{document}